\documentclass[aps,twocolumn,superscriptaddress]{revtex4-1}

\usepackage{amsmath,amssymb,amsfonts,color,graphicx,tabularx}

\usepackage[unicode=true,colorlinks=true]{hyperref}

\hypersetup{linkcolor=blue,citecolor=blue,urlcolor=blue}

\usepackage{mathrsfs}

\usepackage{stackrel}

\usepackage{tikz}

\begin{document}

\title{Resonating valence bond realization of spin-1 non-Abelian chiral spin liquid on the torus}

\author{Hua-Chen Zhang}
\affiliation{Beijing National Laboratory for Condensed Matter Physics \& Institute of Physics, Chinese Academy of Sciences, Beijing 100190, China}
\affiliation{School of Physical Sciences, University of Chinese Academy of Sciences, Beijing 100049, China}

\author{Ying-Hai Wu}
\affiliation{School of Physics and Wuhan National High Magnetic Field Center, Huazhong University of Science and Technology, Wuhan 430074, China}

\author{Hong-Hao Tu}
\email{hong-hao.tu@tu-dresden.de}
\affiliation{Institut f\"ur Theoretische Physik, Technische Universit\"at Dresden, 01062 Dresden, Germany}

\author{Tao Xiang}
\affiliation{Beijing National Laboratory for Condensed Matter Physics \& Institute of Physics, Chinese Academy of Sciences, Beijing 100190, China}
\affiliation{School of Physical Sciences, University of Chinese Academy of Sciences, Beijing 100049, China}

\date{\today}

\begin{abstract}
We propose resonating valence bond wave functions for a spin-1 system on the torus that realize a non-Abelian chiral spin liquid. The wave functions take the form of infinite dimensional matrix product states constructed from conformal blocks of the $\mathrm{SO}(3)_{1}$ Wess-Zumino-Witten model. This means that they are lattice analogues of the bosonic Moore-Read state introduced in fractional quantum Hall systems. The topological order of this system is revealed by explicit construction of three-fold degenerate ground states and analytical computation of the modular S and T matrices.
\end{abstract}

\maketitle

\section{Introduction}

\label{sec:Introduction}

The study of quantum spin liquids is one of the most interesting and challenging topics in current condensed matter physics~\cite{zhou2017,fradkin2013,wen2004,auerbach1994}. These states are disordered ground states of quantum magnets which do not break spin-rotational or lattice symmetries. A famous example is the resonating valence bond (RVB) state constructed by Anderson~\cite{anderson1973}, which was proposed as a candidate for the ground state of the Heisenberg model on the triangular lattice. The term ``valence bond'' (VB) refers to a spin singlet formed by two spin-1/2's on different lattice sites and the coherent superposition of various different VB configurations gives the RVB state. More generally, any singlet-paired state may be defined as a VB state.

After the discovery of the fractional quantum Hall (FQH) effect~\cite{tsui1982,laughlin1983}, Kalmeyer and Laughlin~\cite{kalmeyer1987} realized that the bosonic Laughlin state at half filling can be recast to describe a spin liquid state. This state has spin-rotational symmetry but breaks time-reversal and reflection symmetries, so it was named as a chiral spin liquid (CSL)~\cite{wen1989}. The connection between FQH and CSL states can be understood in the framework of topological order~\cite{wen2004}. When the systems are placed on topologically non-trivial manifolds with nonzero genus (e.g., the torus), there are multiple (quasi-)degenerate ground states that are separated from the rest of the spectrum by an energy gap. The degeneracy is fundamentally determined by the genus of the host manifold and robust against local perturbations. This allows us to describe the low-energy physics using topological quantum field theories (TQFTs). For two-dimensional systems, it has been proposed that universal topological data are encoded in the modular S and T matrices~\cite{keski-vakkuri1993,kitaev2006b}, which can be computed using the degenerate ground states on the torus~\cite{zhang2012}.

One topological property of CSLs is the braiding statistics of their elementary excitations. When two excitations in the Kalmeyer-Laughlin state are braided around each other, the wave function of the system acquires a phase that does not equal $\pm 1$ (as would be the case for bosons or fermions). An even more compelling possibility is the non-Abelian braiding statistics proposed for certain FQH states~\cite{moore1991,wen1991}. In this case, a system with multiple excitations at fixed positions has a few degenerate states, and braiding of two excitations results in a matrix rotation in the degenerate subspace. This property has been extensively pursued for potential applications in topological quantum information processing~\cite{nayak2008}. One prominent example that supports non-Abelian CSL is the Kitaev honeycomb model~\cite{kitaev2006b}. A variety of other non-Abelian CSLs in spin-1 systems have also been proposed~\cite{greiter2009,greiter2014,glasser2015,wildeboer2016,lecheminant2017,liu2018,chen2018}.

The work of Moore and Read~\cite{moore1991} provides another deep insight about chiral topological order from the wave-function perspective as they found that many FQH states can be expressed as conformal blocks, i.e., chiral correlators in certain conformal field theories (CFTs)~\cite{hansson2017}. This connection has been generalized to describe chiral topological states in lattice systems~\cite{cirac2010,nielsen2012,nielsen2014b,tu2014a,tu2014b,bondesan2014,quella2020}, and the wave functions have been formulated as infinite-dimensional matrix product states (IDMPSs). In some cases, the parent Hamiltonians for these states can be derived by using the CFT null field technique~\cite{nielsen2011}.

Besides the CFT approach, the parton method has been used systematically to generate chiral spin liquids and derive their bulk and edge effective theories~\cite{wen1999}. The essential idea of this method is to represent the spins using fermionic or bosonic ``partons'' that reside in enlarged Hilbert spaces. The partons realize the ground states of some mean-field Hamiltonians and physical states of the spins are recovered by applying Gutzwiller projection to remove the unphysical degrees of freedom in the enlarged Hilbert spaces. One extensively used mean-field state of partons is the Bardeen-Cooper-Schrieffer (BCS) state~\cite{bardeen1957a} that describes the spin-singlet pairing of fermions~\cite{baskaran1987,chen2020b}. Along this line, Gutzwiller-projected BCS states provide a convenient way to represent RVB states~\cite{anderson1987}.

In this work, we follow the RVB approach to construct a spin-1 non-Abelian CSL on the torus. The spin-1 system is represented using fermionic partons with three different ``colors''~\cite{liu2010a,liu2012} and each of them forms a projected BCS state with $p+\mathrm{i}p$ pairing. We shall prove that this CSL is the lattice analogue of the bosonic Moore-Read (Pfaffian) state at unit filling. We note that CSLs which exhibit the same topological order as the bosonic Moore-Read state have been extensively studied in the past~\cite{greiter2011,yao2011,tu2013a,greiter2014,glasser2015,wildeboer2016,liu2018}. The important insights revealed in the present work are: (i) By using (generalized) Weierstrass functions as BCS pairing functions, three topologically degenerate ground states, all taking the RVB form, are explicitly constructed on the torus. It is worth emphasizing that the construction is independent of the lattice geometry. (ii) The degenerate ground states have an exact correspondence with the conformal blocks of the $\mathrm{SO}(3)_{1}$ Wess-Zumino-Witten (WZW) model. This generalizes the result of Ref.~\cite{tu2013a}, which studied a single wave function on the plane with open boundaries (and hence was unable to deal with the topological degeneracy). (iii) The degenerate ground states inherit the modular transformation properties of the conformal blocks. For the square lattice, minimally entangled states (MESs)~\cite{zhang2012} are constructed analytically using a proper linear combination of the three degenerate ground states. This allows us to show that the modular S and T matrices coincide with those of the $\mathrm{SO}(3)_{1}$ WZW model, which provides a complete characterization of the non-Abelian topological order of the system.

The rest of this paper is organized as follows. In Sec.~\ref{sec:Wave-function-construction}, we present our construction of the wave functions. In Sec.~\ref{sec:modular}, the topological order of these wave functions is characterized using modular transformations. In Sec.~\ref{sec:Summary-and-discussions}, we summarize our work and give some outlook. In Appendix~\ref{sec:special-functions}, we briefly review the definition and properties of the special functions used in the main text. In Appendix~\ref{sec:cartan}, we introduce the Cartan basis to show that the wave functions can be converted from the Pfaffian form to the ``Jastrow times Pfaffian" form. In Appendix~\ref{sec:translational-invariance}, we prove that the wave functions are translationally invariant on the square lattice.

\section{Wave function construction}

\label{sec:Wave-function-construction}

To begin with, we first briefly review the fermionic parton representation of spin-1 systems~\cite{liu2010a,liu2012}. The discussions in this section are not restricted to any particular lattice geometry and can be applied generally in two-dimensional (2D) lattices. For each lattice site, there is a spin-1 operator obeying the $\mathfrak{so}(3)$ algebra. Its three components are represented using fermionic operators as
\begin{equation}
J_{j,a}=-\mathrm{i}\stackrel[b,c=1]{3}{\sum}\epsilon_{abc}c_{j,b}^{\dagger}c_{j,c},
\label{eq:generator}
\end{equation}
where $\epsilon_{abc}$ is the Levi-Civita symbol and $c_{j,a}^{\dagger}$ ($c_{j,a}$) is the fermionic creation (annihilation) operator at site $j$ ($j=1,\ldots,N$) with color $a$ ($a=1,2,3$). The total number of lattice sites is denoted as $N$ and assumed to be even throughout this paper. The anticommutation relations of the fermionic operators, i.e., $\{c_{i,a},c_{j,b}\} = \{c^\dag_{i,a},c^\dag_{j,b}\} =0$ and $\{c_{i,a},c^\dag_{j,b}\}=\delta_{ij}\delta_{ab}$, help us to confirm that the spin operators satisfy the standard commutation relations
\begin{equation}
[J_{i,a},J_{j,b}]=\delta_{ij}\stackrel[c=1]{3}{\sum}\mathrm{i}\epsilon_{abc}J_{j,c}.
\label{eq:commutator}
\end{equation}
However, the fermionic operators lead to an enlarged Hilbert space compared to the original spin-1 Hilbert space, since only some states in the fermionic Hilbert space are physical spin states. To remove the unphysical states, we impose the single-occupancy constraint $\sum_{a=1}^{3}c_{j,a}^{\dagger}c_{j,a}=1$ at each site. It is also useful to define three ``color'' states
\begin{equation}
\vert a\rangle=c_{a}^{\dagger}\vert0\rangle, \quad a=1,2,3,
\label{eq:localstate}
\end{equation}
where $\vert0\rangle$ is the vacuum of the parton operators and the site index is suppressed for clarity. One can check that the single-site Casimir operator satisfies $\boldsymbol{J}_{j}^{2}\equiv\sum_{a=1}^{3}J_{j,a}^{2}=2$, in agreement with the spin-1 representation. In practice, a state for the spin system is constructed as a many-body state of the fermionic partons, and the single-occupancy constraint is implemented using a Gutzwiller projector $P_{\textrm{G}}$.

\subsection{Projected BCS formulation}

The (unprojected) fermionic parton wave function of our interest takes the BCS form,
\begin{equation}
\vert\Psi_{\textrm{BCS}}\rangle=\exp\left(\underset{i<j}{\sum}g_{ij}\stackrel[a=1]{3}{\sum}c_{i,a}^{\dagger}c_{j,a}^{\dagger}\right)\vert0\rangle,
\label{eq:BCS}
\end{equation}
where $g_{ij}$ is the BCS pairing function and $\sum_{a=1}^{3}c_{i,a}^{\dagger}c_{j,a}^{\dagger}$ is a VB operator creating an $\mathrm{SO}(3)$ singlet between sites $i$ and $j$. The fact that it creates a singlet can be verified by inspecting the eigenvalue of the two-site Casimir operator, $(\boldsymbol{J}_{i}+\boldsymbol{J}_{j})^{2}\sum_{a=1}^{3}c_{i,a}^{\dagger}c_{j,a}^{\dagger}\vert0\rangle=0$. The spin wave function is obtained from Eq.~(\ref{eq:BCS}) as
\begin{eqnarray}
\vert\Psi\rangle &=& P_{\textrm{G}}\vert\Psi_{\textrm{BCS}}\rangle \nonumber \\
                 &=& P_{\textrm{G}}\exp\left(\underset{i<j}{\sum}g_{ij}\stackrel[a=1]{3}{\sum}c_{i,a}^{\dagger}c_{j,a}^{\dagger}\right)\vert0\rangle,
\label{eq:projectedBCS}
\end{eqnarray}
where the Gutzwiller projector $P_{\textrm{G}}$ removes all non-singly occupied states. When the exponential is expanded, a coherent superposition of VB singlet configurations survives in the Gutzwiller projection, so the spin wave function $\vert\Psi\rangle$ in Eq.~(\ref{eq:projectedBCS}) is an RVB state.

In Ref.~\cite{tu2013a}, it was revealed that the particular choice $g_{ij} = 1/(z_i-z_j)$ of the pairing function in Eq.~(\ref{eq:projectedBCS}), with $z_j$ being the complex coordinate of site $j$ on the plane, gives rise to a lattice version of the bosonic Moore-Read state at unit filling. This pairing function describes a $p+\mathrm{i}p$ superconductor in its weak-pairing topological phase and coincides with the two-point correlator of a chiral Majorana field in the Ising CFT~\cite{read2000}. This observation serves as a guide to recast the projected BCS state in Eq.~(\ref{eq:projectedBCS}) as a multipoint correlator of three-colored Majorana fields or, in other words, an IDMPS constructed from the $\mathrm{SO}(3)_{1}$ WZW model, where the latter CFT describes the gapless edge excitations of the bosonic Moore-Read state. That is, the wave function for the bulk topological order and the dynamical theory at the edge are described by the same CFT, in agreement with the bulk-edge correspondence~\cite{moore1991}.

The results of Ref.~\cite{tu2013a} were obtained for an infinite plane, which means that the projected BCS state in Eq.~(\ref{eq:projectedBCS}) describes a \emph{unique} ground state on a 2D lattice with \emph{open} boundaries. On the other hand, many interesting and important properties of topologically ordered systems, such as the ground-state degeneracy and modular matrices, cannot be revealed on a topologically trivial manifold like the open plane. It is thus desirable to generalize the wave function to topologically nontrivial manifolds, the simplest of which is a torus with \emph{periodic} boundaries along both directions of a 2D lattice. From a numerical perspective, the torus geometry suppresses boundary effects and allows for the usage of translation symmetries, so it is possible to study larger systems, which might be helpful in the search and identification of bosonic Moore-Read states in lattice models~\cite{glasser2015}.

To construct projected BCS states in periodic systems, it is natural to expect that the pairing function $g_{ij}$ in Eq.~(\ref{eq:projectedBCS}) should be replaced by the two-point correlator of Majorana fields in the Ising CFT on the torus. In this case, the Majorana correlator $\langle\chi(z_{i})\chi(z_{j})\rangle$ becomes the elliptic generalization of $1/(z_{i}-z_{j})$. However, there is an extra (and crucial) complication for the torus geometry: Majorana fermions in the Ising CFT could have either periodic ($\textrm{P}$) or antiperiodic ($\textrm{A}$) boundary condition along the two incontractible large loops of the torus~\cite{ginsparg1988,francesco1997}. Consequently, there are four topological sectors $\textrm{PP, PA, AA, AP}$, which hereafter shall be labelled as $\nu=1,2,3,4$, respectively.

\begin{figure}
\includegraphics[width=\linewidth]{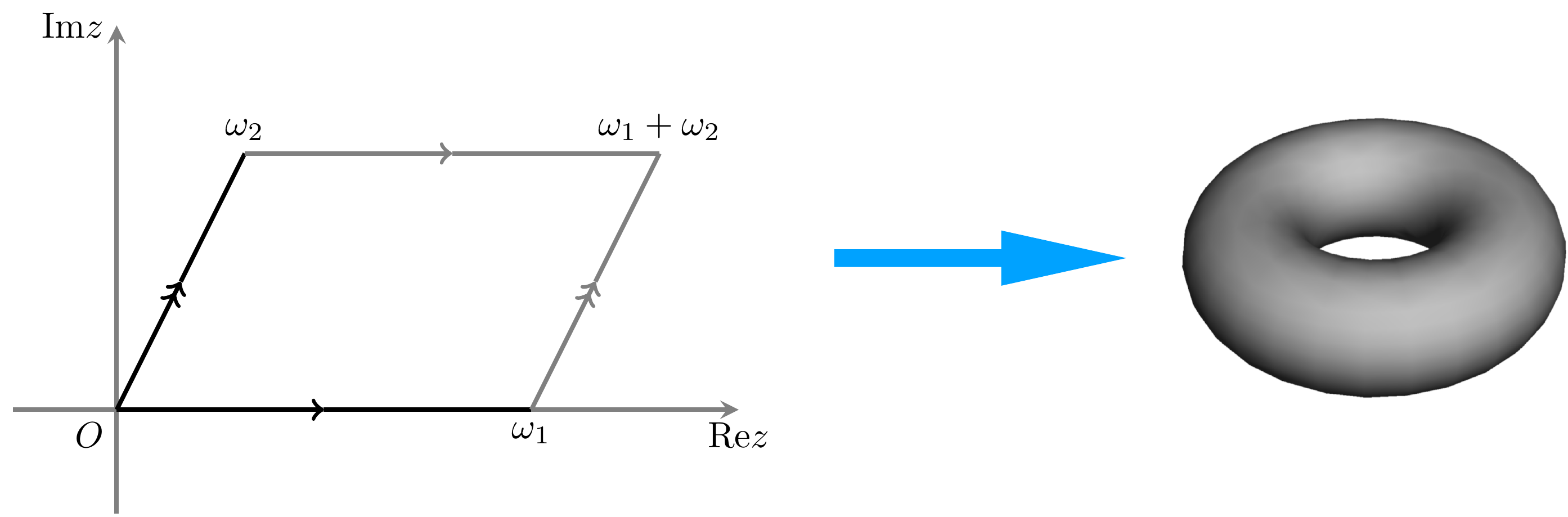}
\caption{Definition of the torus by specifying two (noncollinear) complex numbers $\omega_{1}$ and $\omega_{2}$ in the complex plane, where $\omega_{1}$ is chosen to be real and positive. $\omega_{1}$ and $\omega_{2}$ define a parallelogram and the torus is obtained by identifying the opposite sides of the parallelogram. The modular parameter $\tau$ is the aspect ratio of the parallelogram, $\tau=\omega_{2}/\omega_{1}$. In the present work, we choose $\mathrm{Im}\tau>0$.}
\label{fig:torus}
\end{figure}

For the torus with modular parameter $\tau$ (see Fig.~\ref{fig:torus}), the two-point correlator of the chiral Majorana field $\chi(z)$ vanishes in the $\nu=1$ sector due to the presence of a zero mode and is given by
\begin{equation}
\langle\chi(z_{i})\chi(z_{j})\rangle_{\nu}=\mathscr{P}_{\nu}(z_{i}-z_{j}\vert\tau)
\label{eq:2-point-correlator}
\end{equation}
in the $\nu=2,3,4$ sectors~\cite{francesco1987,francesco1997}, where $\mathscr{P}_{\nu}(z_{i}-z_{j}\vert\tau)$ are (generalized) Weierstrass functions [see Eq.~(\ref{eq:Weierstrass}) in Appendix~\ref{sec:special-functions} for the definition]. With the (generalized) Weierstrass functions as pairing functions, \emph{three} projected BCS states on the torus can be constructed as
\begin{equation}
\vert\Psi_{\nu}\rangle=P_{\textrm{G}}\exp\left(\underset{i<j}{\sum}\mathscr{P}_{\nu}(z_{i}-z_{j}\vert\tau)\stackrel[a=1]{3}{\sum}c_{i,a}^{\dagger}c_{j,a}^{\dagger}\right)\vert0\rangle.
\label{eq:RVB}
\end{equation}
As we shall elaborate in subsequent sections, these are indeed natural generalizations of the projected BCS state on the plane studied in Ref.~\cite{tu2013a}. These states span the topologically degenerate ground-state subspace and exhibit the SO(3)$_1$ topological order.

\subsection{Conformal field theory formulation}

In Ref.~\cite{tu2013a}, much insight about the projected BCS state on the plane was derived from its exact equivalence with an IDMPS constructed from the SO(3)$_1$ WZW model. It is very interesting that this equivalence still holds on the torus. In other words, we shall prove that the three projected BCS states in Eq.~(\ref{eq:RVB}) can also be formulated as conformal blocks of the SO(3)$_1$ WZW model on the torus.

As a bridge between the two formulations, we first derive the explicit form of the wave functions in the spin basis. By expanding the BCS states (in terms of Pfaffians in real space) and performing the Gutzwiller projection, Eq.~(\ref{eq:RVB}) is transformed to
\begin{equation}
\vert\Psi_{\nu}\rangle = \sum_{a_1,\ldots,a_N} \Psi_{\nu}(a_{1},\ldots,a_{N}) \vert a_{1},\ldots,a_{N}\rangle,
\end{equation}
where the coefficients are, up to a sign factor, a product of three Pfaffians,
\begin{eqnarray}
\Psi_{\nu}(a_{1},\ldots,a_{N}) &=& \mathrm{sgn}(i_{1}^{(1)},\ldots,i_{N_1}^{(1)},\ldots,i_1^{(3)},\ldots,i_{N_{3}}^{(3)}) \nonumber \\
&\phantom{=}&\times \stackrel[a=1]{3}{\prod}\mathrm{Pf}_{a}\left[\mathscr{P}_{\nu}(z_{i}-z_{j}\vert\tau)\right],
\label{eq:IDMPS_from_RVB}
\end{eqnarray}
and the basis states can be represented using the fermionic parton operators as
\begin{equation}
\vert a_{1},\ldots,a_{N}\rangle=c_{1,a_{1}}^{\dagger}\cdots c_{N,a_{N}}^{\dagger}\vert 0\rangle.
\label{eq:spin_basis}
\end{equation}
In Eq.~(\ref{eq:IDMPS_from_RVB}), the lattice sites occupied by $\vert a\rangle$ are denoted by $i_1^{(a)}<\cdots<i_{N_a}^{(a)}$. We note that only \emph{even} $N_a$ is allowed in Eq.~(\ref{eq:IDMPS_from_RVB}) because it is implicit in Eq.~(\ref{eq:RVB}) that the partons of each color must come in pairs. Regarding Eq.~(\ref{eq:IDMPS_from_RVB}), two additional comments are in order: (i) The factor $\mathrm{sgn}(i_{1}^{(1)},\ldots,i_{N_{1}}^{(1)},\ldots,i_{1}^{(3)},\ldots,i_{N_{3}}^{(3)})=\pm1$ is the signature of the permutation which brings $(i_{1}^{(1)},\ldots,i_{N_{1}}^{(1)},\ldots,i_{1}^{(3)},\ldots,i_{N_{3}}^{(3)})$ back to $(1,\ldots,N)$. (ii) $\mathrm{Pf}_{a}\left[\mathscr{P}_{\nu}(z_{i}-z_{j}\vert\tau)\right]$
is the Pfaffian of an $N_{a}\times N_{a}$ antisymmetric matrix, whose diagonal entries are zero and off-diagonal entries are $\mathscr{P}_{\nu}(z_{i}-z_{j}\vert\tau)$, in which the complex coordinates are restricted to the positions of the spin state $\vert a\rangle$.

Let us now turn to the CFT side and interpret the wave functions as conformal blocks of the $\mathrm{SO}(3)_{1}$ WZW model. This CFT has central charge $c=3/2$ and can be formulated as a \emph{free} theory of three copies of massless Majorana fermion, which has a natural correspondence with the Ising CFT. For instance, we may identify its Kac-Moody primary field in the vector representation with a (three-colored) energy operator of the Ising CFT,
\begin{equation}
\varepsilon^{a}(z,\bar{z})=\mathrm{i}\chi^{a}(z)\bar{\chi}^{a}(\bar{z}), \quad a=1,2,3,
\label{eq:energy_operator}
\end{equation}
where the holomorphic component $\chi^{a}(z)$ is a chiral Majorana field with color $a$, and $\bar{\chi}^{a}(\bar{z})$ is its antiholomorphic counterpart. For energy operators with the same color, the multipoint correlator on the torus breaks up into a sum of the products of holomorphic and antiholomorphic components as follows~\cite{francesco1997}:
\begin{equation}
\langle\varepsilon^{a}(z_{1},\bar{z}_{1})\varepsilon^{a}(z_{2},\bar{z}_{2})\cdots\rangle^{\prime}=\underset{\nu}{\sum}~\vert\langle\chi^{a}(z_{1})\chi^{a}(z_{2})\cdots\rangle_{\nu}^{\prime}\vert^{2},
\label{eq:energy_correlator}
\end{equation}
where $\langle\cdots\rangle^{\prime}$ stands for the correlator without division of the corresponding partition function (i.e., ``unnormalized'' in the field theory sense) and $\nu$ labels the four boundary conditions of the Majorana fermions on the torus. If the number of energy operators in Eq.~(\ref{eq:energy_correlator}) is even, nontrivial contributions come from the $\nu=2,3,4$ sectors (PA, AA, AP boundary conditions, respectively), while such correlator vanishes in the $\nu=1$ sector (PP boundary condition).

We are now in a position to write down the following wave functions using conformal blocks on the torus~\cite{nielsen2014b,deshpande2016}:
\begin{align}
\vert\psi_{\nu}\rangle &= \sum_{a_1,\ldots,a_N} \langle\chi^{a_{1}}(z_{1})\cdots\chi^{a_{N}}(z_{N})\rangle_{\nu}^{\prime} \vert a_{1},\ldots,a_{N}\rangle \nonumber \\
&\equiv \sum_{a_1,\ldots,a_N} \psi_{\nu}(a_{1},\ldots,a_{N}) \vert a_{1},\ldots,a_{N}\rangle,
\label{eq:IDMPS_from_CFT}
\end{align}
where $\nu=2,3,4$~\footnote{For $\nu=1$ and even $N$, one can prove that the conformal block vanishes. In that case, grouping together Majorana fields (like Eq.~(\ref{eq:wave_function_from_CFT})) inevitably yields a correlator with even number of Majorana fields with the same color, which vanishes for $\nu=1$ (PP boundary condition).}. Such wave functions are often termed as ``infinite-dimensional matrix product states''~\cite{cirac2010,nielsen2012,tu2015} due to the formal analogy to usual matrix product states (MPSs), for which the wave-function coefficients are written as a product of local matrices associated with each lattice site. For IDMPSs, the finite-dimensional matrices defining the MPSs are replaced by a set of fields living in the ``infinite-dimensional'' Hilbert space of CFTs.

At this stage, a crucial observation is that the wave functions defined in Eq.~(\ref{eq:IDMPS_from_CFT}) are identical to those defined by Eq.~(\ref{eq:IDMPS_from_RVB}).
This can be proved as
\begin{eqnarray}
\psi_{\nu}(a_1,\ldots,a_N)
&=& \zeta\stackrel[a=1]{3}{\prod}\langle\chi^{a}(z_{i_{1}^{(a)}})\cdots\chi^{a}(z_{i_{N_{a}}^{(a)}})\rangle_{\nu}^{\prime} \nonumber \\
&=& \zeta Z^3_{\nu}\stackrel[a=1]{3}{\prod}\langle\chi^{a}(z_{i_{1}^{(a)}})\cdots\chi^{a}(z_{i_{N_{a}}^{(a)}})\rangle_{\nu} \nonumber \\
&=& Z^3_{\nu} \zeta \stackrel[a=1]{3}{\prod}\mathrm{Pf}_{a}\left[\mathscr{P}_{\nu}(z_{i}-z_{j}\vert\tau)\right] \nonumber \\
&=& Z^3_{\nu} \Psi_{\nu}(a_{1},\ldots,a_{N}).
\label{eq:wave_function_from_CFT}
\end{eqnarray}
In the first line, the Majorana fields with the same color are grouped together, where the permutation sign $\zeta=\mathrm{sgn}(i_{1}^{(1)},\ldots,i_{N_{1}}^{(1)},\ldots,i_{1}^{(3)},\ldots,i_{N_{3}}^{(3)})$ arising due to the anticommuting nature of Majorana fields is exactly the same as the sign factor in Eq.~(\ref{eq:IDMPS_from_RVB}). In the second line, the unnormalized correlator is converted to a normalized one, by taking into account the partition function of three massless chiral Majorana fermions in the respective sector. The torus partition function of a \emph{single} massless chiral Majorana fermion in the sector $\nu$ is denoted as $Z_{\nu}$. It is worth emphasizing that the three Majorana fermions are locked in the \emph{same} sector, as required in the SO(3)$_1$ WZW model. In the third line, Wick's theorem~\cite{francesco1997} is used to reduce the (normalized) multipoint correlators to two-point ones in Eq.~(\ref{eq:2-point-correlator}). At this stage, we also see that $N_a$ (i.e., the number of sites with color $a$ in a given spin configuration) must be even, since the correlator of an odd number of Majorana fields vanishes in the $\nu=2,3,4$ sectors~\cite{francesco1997}.

We have thus established that the IDMPSs constructed in Eq.~(\ref{eq:IDMPS_from_CFT}) are identical to the wave functions in Eq.~(\ref{eq:IDMPS_from_RVB}), up to overall factors $Z_{\nu}^3$. If the IDMPSs were defined with ``normalized'' correlators with the partition functions included properly, the overall factors would have disappeared. However, as we shall see in Sec.~\ref{sec:modular}, keeping these factors is useful for understanding how these wave functions change under modular transformations.

Finally, as a side remark, we note that the wave functions in Eq.~(\ref{eq:IDMPS_from_RVB}), when converted to the Cartan basis, exhibit a ``Jastrow times Pfaffian" form [see Eq.~(\ref{eq:jastrow_pfaffian})], in complete analogy to their planar counterpart~\cite{nielsen2011,tu2013a}. The Cartan basis for which one of the $\mathfrak{so}(3)$ generators [Eq.~(\ref{eq:generator})] is diagonal is more suitable for numerical calculations. The detailed derivation of this ``Jastrow times Pfaffian" form is provided in Appendix~\ref{sec:cartan}.

\section{Modular S and T transformations of the wave functions}

\label{sec:modular}

The periodic structure of the torus allows for different parametrizations. The reparametrizations that leave the torus invariant are modular transformations~\cite{francesco1997,ginsparg1988,blumenhagen2009}. In the continuum, all modular transformations can be generated by two elementary ones, namely, the S and T transformations. For topological orders in two dimensions, the modular S and T transformations act in the degenerate ground-state subspace and their matrix forms encode useful information about the underlying anyon theory~\cite{keski-vakkuri1993,kitaev2006b}.

\begin{figure}
\includegraphics[width=\linewidth]{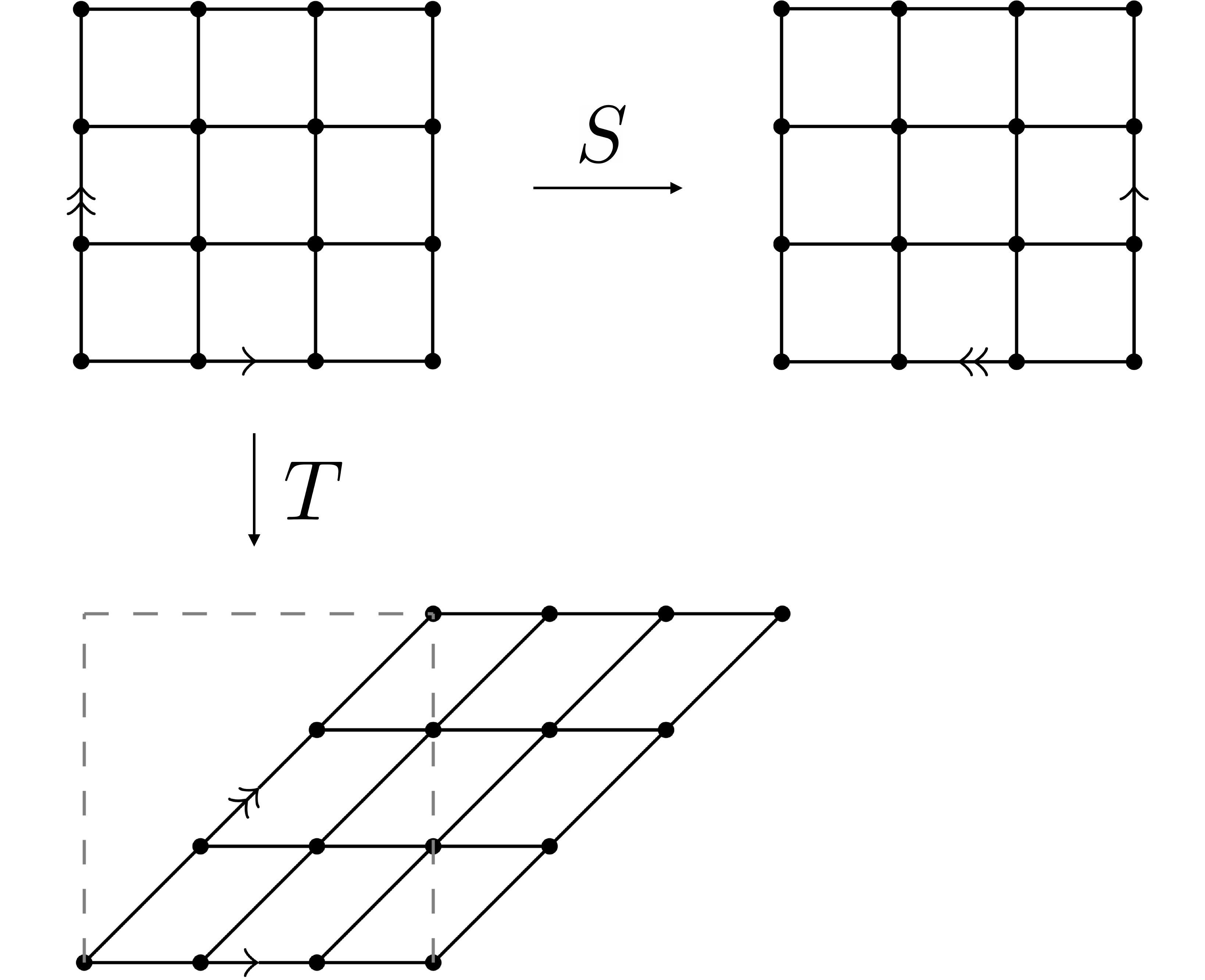}
\caption{Schematics of the modular S and T transformations on a $4 \times 4$ square lattice. The lattice is periodic along both directions. The S transformation corresponds to a 90-degree counterclockwise rotation, whilst the T transformation corresponds to a Dehn twist. The arrow and double-arrow indicate the orientation of the lattice.}
\label{fig:ModularTransformation}
\end{figure}

The presence of a lattice further constrains allowed modular transformations, since the lattice should coincide with itself after the modular transformations~\cite{zhang2012}. To calculate how the wave functions change under the modular transformations, we consider an $L \times L$ square lattice for simplicity, where $L$ is taken to be even to ensure that $N=L^2$ is even. Without loss of generality, we assume that the lattice spacing is $1/L$, so that the lattice is embedded on a square with a unit side length ($\omega_1=1$ and $\omega_2=\mathrm{i}$, see Fig.~\ref{fig:torus}) and its modular parameter is $\tau=\mathrm{i}$. For site $j$ with complex coordinate $z_{j} = \tfrac{1}{L}(x_j + \mathrm{i} y_j)$, we choose $x_{j}\in\{0,1,\ldots,L-1\}$ and $y_{j}\in\{0,1,\ldots ,L-1\}$ such that
\begin{equation}
\label{eq:labelling}
j=x_{j}+y_{j}L+1\equiv(x_{j},y_{j}).
\end{equation}

To determine transforming properties of the wave functions under the modular transformations, we consider a process in which the modular parameter $\tau$ of the torus changes adiabatically. The implementation of the modular transformation is then an adiabatic transport along a closed path in the ``moduli space'' of distinct tori, which is obtained from the upper half plane (recall our assumption $\mathrm{Im}\tau>0$) by identifying the $\tau$-points that are related by modular transformations.
We shall follow the ``holonomy equals monodromy'' approach~\cite{read2009}. Here ``holonomy" refers to the (non-Abelian) gauge-invariant Berry phase of the wave functions under the adiabatic transport, whilst ``monodromy" is just analytical continuation of the positions of primary fields in the conformal blocks. In Ref.~\cite{read2009}, ``holonomy equals monodromy'' was demonstrated for the continuous case by considering special paths in the ``moduli space''. As our lattice wave functions should have short-range (exponentially decaying) correlations, it is expected that ``holonomy equals monodromy'' holds when $L$ is much larger than the correlation length.

We are now ready to calculate how the wave functions $\vert \psi_{\nu}\rangle$ in Eq.~\eqref{eq:wave_function_from_CFT} change under the modular transformations. As we noted earlier, the ``normalization'' factor $Z_{\nu}$ is the chiral partition function of a massless free Majorana fermion in the $\nu=2,3,4$ sectors on the torus. Their explicit forms are given by~\cite{francesco1997}
\begin{align}
Z_{2}(\tau)&=\sqrt{\frac{\vartheta_{2}(\tau)}{\eta(\tau)}}=\sqrt{2}q^{\frac{1}{24}}\stackrel[n=1]{\infty}{\prod}(1+q^{n}), \nonumber \\
Z_{3}(\tau)&=\sqrt{\frac{\vartheta_{3}(\tau)}{\eta(\tau)}}=q^{-\frac{1}{48}}\stackrel[n=0]{\infty}{\prod}(1+q^{n+\frac{1}{2}}), \nonumber \\
Z_{4}(\tau)&=\sqrt{\frac{\vartheta_{4}(\tau)}{\eta(\tau)}}=q^{-\frac{1}{48}}\stackrel[n=0]{\infty}{\prod}(1-q^{n+\frac{1}{2}}),
\label{eqs:normalisation}
\end{align}
where $q=\mathrm{e}^{2\pi\mathrm{i}\tau}$, $\vartheta_{\nu}(\tau)$ are Jacobi's theta functions (see Appendix~\ref{sec:special-functions}) and
\begin{equation}
\eta(\tau)=q^{\frac{1}{24}}\prod_{n=1}^{\infty}(1-q^{n})
\label{eq:dedekind_eta}
\end{equation}
is Dedekind's eta function. It should be emphasized that these factors are $\tau$-dependent, which is known~\cite{read2009} to be important in understanding the modular transformation properties of the wave functions.

\subsection{Modular T transformation}

\label{subsec:modular_T}

Geometrically, the modular T transformation on the torus corresponds to the so-called Dehn twist (see Fig.~\ref{fig:ModularTransformation}), the action of which on the coordinates of lattice sites is $(x_{j},y_{j})\rightarrow(x_{j}+y_{j},y_{j})$, in which the addition is modulo $L$. Here we again follow the aforementioned ``holonomy equals monodromy'' approach, which changes the modular parameter of the torus as $\tau\rightarrow\tau+1$.

Under the change $\tau\rightarrow\tau+1$, the factors in Eq.~(\ref{eqs:normalisation}) transform as
\begin{align}
Z_{2}(\tau+1)&=\mathrm{e}^{\frac{\mathrm{i}\pi}{12}}Z_{2}(\tau), \nonumber \\
Z_{3}(\tau+1)&=\mathrm{e}^{-\frac{\mathrm{i}\pi}{24}}Z_{4}(\tau), \nonumber \\
Z_{4}(\tau+1)&=\mathrm{e}^{-\frac{\mathrm{i}\pi}{24}}Z_{3}(\tau),
\label{eqs:normalisation_T}
\end{align}
which are easily obtained by using the modular transformation properties of Jacobi's theta functions~(\ref{eqs:theta_T}) and Dedekind's eta function~(\ref{eq:eta_T}).

By using Eqs.~\eqref{eqs:normalisation_T} and~\eqref{eqs:Weierstrass_T}, we find that after the Dehn twist, the wave functions $\vert \psi_{\nu}\rangle$ defined in Eq.~(\ref{eq:wave_function_from_CFT}) are transformed to
\begin{align}
\vert\psi'_2\rangle&=\mathrm{e}^{\frac{\mathrm{i}\pi}{4}}\vert\psi_2\rangle, \nonumber \\
\vert\psi'_3\rangle&=\mathrm{e}^{-\frac{\mathrm{i}\pi}{8}}\vert\psi_4\rangle, \nonumber \\
\vert\psi'_4\rangle&=\mathrm{e}^{-\frac{\mathrm{i}\pi}{8}}\vert\psi_3\rangle.
\end{align}
This motivates us to define the so-called minimally entangled states (MESs)~\cite{zhang2012}, also known as the anyon eigenstates, as
\begin{eqnarray}
\vert\psi_\mathrm{I}\rangle &=& \frac{1}{\sqrt{2}}(\vert\psi_3\rangle+\vert\psi_4\rangle), \nonumber \\
\vert\psi_\mathrm{s}\rangle &=& \vert\psi_2\rangle, \nonumber \\
\vert\psi_\mathrm{v}\rangle &=& \frac{1}{\sqrt{2}}(\vert\psi_3\rangle-\vert\psi_4\rangle).
\label{eq:MES}
\end{eqnarray}
Under the Dehn twist, the MESs are transformed to themselves up to a phase factor
\begin{eqnarray}
\vert\psi'_\mathrm{I}\rangle&=\mathrm{e}^{-\frac{\mathrm{i}\pi}{8}}\vert\psi_\mathrm{I}\rangle, \nonumber\\
\vert\psi'_\mathrm{s}\rangle&=\mathrm{e}^{\frac{\mathrm{i}\pi}{4}}\vert\psi_\mathrm{s}\rangle, \nonumber \\
\vert\psi'_\mathrm{v}\rangle&=-\mathrm{e}^{-\frac{\mathrm{i}\pi}{8}}\vert\psi_\mathrm{v}\rangle.
\label{eq:MES_Dehntwist}
\end{eqnarray}
In other words, the modular T matrix, which is defined as the transforming matrix relating the states before and after the Dehn twist, is diagonal within the MES basis,
\begin{equation}
T=\left(\begin{array}{ccc}
\mathrm{e}^{-\frac{\mathrm{i}\pi}{8}} & 0 & 0\\
0 & \mathrm{e}^{\frac{\mathrm{i}\pi}{4}} & 0\\
0 & 0 & -\mathrm{e}^{-\frac{\mathrm{i}\pi}{8}}
\end{array}\right).
\label{eq:T}
\end{equation}

The diagonal entries of the $T$ matrix are known to encode the topological spins of anyonic quasiparticles as well as the chiral central charge via $T_{\alpha \alpha} = \theta_{\alpha}\mathrm{e}^{-2\pi\mathrm{i}c/24}$, where $\alpha$ denotes the anyon type ($\alpha=\mathrm{I},\mathrm{s},\mathrm{v}$), $\theta_{\alpha}$ are their topological spins and $c$ is the chiral central charge of the edge CFT of the bulk anyon theory. By using Eq.~(\ref{eq:T}), we obtain $\theta_{\mathrm{I}}=1$, $\theta_{\mathrm{s}}=\mathrm{e}^{2\mathrm{i}\pi\frac{3}{16}}$, $\theta_{\mathrm{v}}=-1$, and $c=3/2$ (mod 8). This is in agreement with the SO(3)$_1$ TQFT~\cite{kitaev2006b}, which indeed describes the topological order of the bosonic Moore-Read state at unit filling. From the CFT side, the MESs are in correspondence with the primary fields, and we could extract the conformal weights of these primary fields by using $\theta_{\alpha} = \mathrm{e}^{2\pi\mathrm{i}h_{\alpha}}$, where $h_{\alpha}$ is the conformal weight of the primary field corresponding to $\alpha$. For our case, this gives rise to $h_{\mathrm{I}}=0$, $h_{\mathrm{s}}=\frac{3}{16}$, $h_{\mathrm{v}}=\frac{1}{2}$, which correspond to the three primary fields of the SO(3)$_{1}$ WZW model.

\subsection{Modular S transformation}

Geometrically, the modular S transformation is a counterclockwise rotation of the lattice by 90 degrees [as shown in Fig.~\ref{fig:ModularTransformation}], the action of which on the coordinates of lattice sites is $(x_{j},y_{j})\rightarrow(y_{j},L-1-x_{j})$ or $z_{j}\rightarrow-\mathrm{i}z_{j}$. Equivalently, we write it in the form of a map $d$~\cite{nielsen2014b},
\begin{equation}
d(x_{j}+y_{j}L+1)=y_{j}+(L-1-x_{j})L+1.
\label{eq:relabelling}
\end{equation}
This transformation is realized via $\tau\rightarrow-1/\tau$. Again, we follow the ``holonomy equals monodromy'' approach.

Under the change $\tau\rightarrow-1/\tau$, the factors in Eq.~(\ref{eqs:normalisation}) transform as
\begin{align}
Z_{2}(-1/\tau)&=Z_{4}(\tau), \nonumber \\
Z_{3}(-1/\tau)&=Z_{3}(\tau), \nonumber \\
Z_{4}(-1/\tau)&=Z_{2}(\tau),
\label{eqs:normalisation_S}
\end{align}
which come from the modular transformation properties of Jacobi's theta functions~(\ref{eq:theta_S_1})-(\ref{eq:theta_S_4}) and Dedekind's eta function~(\ref{eq:eta_S}).

By combining Eqs.~(\ref{eqs:normalisation_S}) and (\ref{eqs:Weierstrass_S}), we find that under the modular S transformation, the MESs $\{\vert\psi_{\mathrm{I}}\rangle,\vert\psi_{\mathrm{s}}\rangle,\vert\psi_{\mathrm{v}}\rangle\}$ defined in Eq.~(\ref{eq:MES}) are transformed to
\begin{align}
\vert\psi''_{\mathrm{I}}\rangle&=\frac{1}{2}\vert\psi_{\mathrm{I}}\rangle+\frac{1}{\sqrt{2}}\vert\psi_{\mathrm{s}}\rangle+\frac{1}{2}\vert\psi_{\mathrm{v}}\rangle, \nonumber \\
\vert\psi''_{\mathrm{s}}\rangle&=\frac{1}{\sqrt{2}}(\vert\psi_{\mathrm{I}}\rangle-\vert\psi_{\mathrm{v}}\rangle), \nonumber \\
\vert\psi''_{\mathrm{v}}\rangle&=\frac{1}{2}\vert\psi_{\mathrm{I}}\rangle-\frac{1}{\sqrt{2}}\vert\psi_{\mathrm{s}}\rangle+\frac{1}{2}\vert\psi_{\mathrm{v}}\rangle.
\end{align}
Thus, the modular S matrix is written as
\begin{equation}
S=\frac{1}{2}\left(\begin{array}{ccc}
1 & \sqrt{2} & 1\\
\sqrt{2} & 0 & -\sqrt{2}\\
1 & -\sqrt{2} & 1
\end{array}\right),
\label{eq:S}
\end{equation}
which is indeed the expected form for the $\mathrm{SO}(3)_{1}$ TQFT~\cite{kitaev2006b}.

As a technical detail, we note that the map~\eqref{eq:relabelling} is a relabelling of the sites in the lattice. It turns out that this map contains $N(N-2L+1)/4$ exchanges of two labels. The resulting extra sign factor $(-1)^{N(N-2L+1)/4}$ exactly cancels the factor $\mathrm{i}^{N/2}$ coming from the transformation of (generalized) Weierstrass functions [Eq.~\eqref{eqs:Weierstrass_S}], since we have $N=L^{2}$ and $N$ is even.

\subsection{Orthogonality}

\label{sec:Orthogonality}

In the thermodynamic limit, we expect that the MES basis states are orthogonal to each other. This is quantified by the wave-function overlaps encoded in the Gram matrix $G_{\alpha\beta} = \langle \psi_{\alpha} \vert \psi_{\beta} \rangle$ ($\alpha, \beta=\mathrm{I},\mathrm{v},\mathrm{s}$), which should behave as $G_{\alpha\beta} \propto \delta_{\alpha\beta}$ for large $L$. Unlike the modular matrices, these wave-function overlaps do not seem to have a closed form and, thus, an exact computation is limited to small system sizes. For a $4 \times 4$ square lattice, the wave-function overlaps for normalized MES basis states are found to be $|G_{\mathrm{I},\mathrm{v}}| = 0.03609$, $|G_{\mathrm{I},\mathrm{s}}| = 0.26608$, and $|G_{\mathrm{v},\mathrm{s}}| = 0.26342$, which are already small. For larger system sizes, the verification of the vanishing wave-function overlaps can be done with Monte Carlo techniques~\cite{glasser2015,wildeboer2016,liu2018}. These more expensive numerical calculations are left for future work.

\section{Summary and discussions}

\label{sec:Summary-and-discussions}

To summarize, we have constructed three RVB wave functions of a spin-1 system that represent the topologically degenerate ground states of a bosonic Moore-Read-type chiral spin liquid on the torus. These wave functions take the form of projected BCS states and can also be formulated as conformal blocks of the $\mathrm{SO}(3)_{1}$ WZW model. For the case of a square lattice, we have built the minimally entangled state basis from the three RVB wave functions and analytically computed modular S and T matrices within this basis. The results are in agreement with the SO(3)$_1$ topological quantum field theory and thus fully characterize the topological order.

For Moore-Read-type chiral spin liquid states realized in spin-1 systems, short-range Hamiltonians supporting them have so far not been settled, despite that a number of candidates were suggested~\cite{greiter2009,glasser2015,lecheminant2017,liu2018,chen2018}. To find suitable Hamiltonians, the degenerate ground states constructed in the present work could be used for optimizing the overlaps with the numerically obtained ground states of candidate Hamiltonians. It could help constrain the parameter space of the candidate Hamiltonians. In parallel, it is also desirable to develop the parent Hamiltonian formalism for conformal block wave functions on the torus. Unlike the planar case, such formalism is not yet available for the torus case~\cite{sierra2020}.

The spin-1 RVB wave functions considered in this work have a natural SO($n$) generalization (see Ref.~\cite{tu2013a} for the planar case). This provides wave-function realizations of the 16 types of topological orders known as Kitaev's sixteenfold way~\cite{kitaev2006b}, which complements the Hamiltonian construction~\cite{chulliparambil2020}. For odd $n$, the three states similar to Eq.~\eqref{eq:RVB} (with the VB creation operator being replaced by its SO($n$) analogue) span the three-dimensional ground-state subspace, which belongs to the non-Abelian series of Kitaev's sixteenfold way. However, for the Abelian series with even $n$, the ground-state subspace should be four-dimensional and the construction similar to Eq.~\eqref{eq:RVB} would miss one state. This remains a puzzle and it would be interesting to find out this missing state in future works.

Finally, it is worth noting that the study of chiral spin liquids is not a topic of only theoretical interest. The recent experimental observation of the half-quantized thermal Hall conductivity in the Kitaev material $\alpha$-$\textrm{RuCl}_{3}$ hints at the emergence of a non-Abelian chiral spin liquid~\cite{kasahara2018}. For certain materials, a spin-1 chiral spin liquid (with $p+\mathrm{i}p$ pairing of fermionic spinons) has also been suggested as a candidate ground state~\cite{liu2010a,liu2010b,serbyn2011,bieri2012}. This state has the same topological order as the RVB wave functions considered here, but it remains elusive whether such states can appear as ground states of a simple lattice model that is relevant for materials or realizable in quantum simulators. It is our hope that the ``idealized'' wave functions constructed in the present work could help in the search for suitable lattice models, which might in turn shed light on possible experimental realizations.

\acknowledgments

We are grateful to Anne Nielsen and Germ\'an Sierra for helpful discussions. H.H.T. acknowledges Zheng-Xin Liu, Yi Zhou, and Tai-Kai Ng for collaborations on related topics. The authors are supported by the National Key Research and Development Project of China under Grant No.~2017YFA0302901, the National Natural Science Foundation of China under Grants No.~11888101, No.~11804107, startup grant of HUST, and the DFG through project A06 of SFB 1143 (project-id 247310070).

\appendix

\section{Jacobi's theta functions and generalized Weierstrass functions}

\label{sec:special-functions}

In this Appendix, we briefly review the definition and some properties of Jacobi's theta functions and (generalized) Weierstrass functions, which are necessary for the derivation in this work. For further details about these functions, one can refer to Refs.~\cite{francesco1997,blumenhagen2009}.

\subsection{Definitions}

Jacobi's theta functions are defined by
\begin{align}
\vartheta_{1}(z\vert\tau)&=-\mathrm{i}\underset{n\in\mathbb{Z}}{\sum}(-1)^{n}y^{n+1/2}q^{(n+1/2)^{2}/2}, \nonumber \\
\vartheta_{2}(z\vert\tau)&=\underset{n\in\mathbb{Z}}{\sum}y^{n+1/2}q^{(n+1/2)^{2}/2}, \nonumber \\
\vartheta_{3}(z\vert\tau)&=\underset{n\in\mathbb{Z}}{\sum}y^{n}q^{n^{2}/2}, \nonumber \\
\vartheta_{4}(z\vert\tau)&=\underset{n\in\mathbb{Z}}{\sum}(-1)^{n}y^{n}q^{n^{2}/2},
\label{Jacobi_theta}
\end{align}
where $q=\mathrm{e}^{2\pi\mathrm{i}\tau}$ and $y=\mathrm{e}^{2\pi\mathrm{i}z}$. The argument of these functions is the complex variable $z$, whilst $\tau$ is a complex parameter with $\mathrm{Im}\tau>0$. By using Jacobi's triple product identity~\cite{francesco1997,ginsparg1988,blumenhagen2009}
\begin{equation}
\stackrel[n=1]{\infty}{\prod}(1-q^{n})(1+yq^{n-1/2})(1+y^{-1}q^{n-1/2})=\underset{n\in\mathbb{Z}}{\sum}y^{n}q^{n^{2}/2},
\end{equation}
Jacobi's theta functions can also be expressed in the form of infinite products:
\begin{align}
\vartheta_{1}(z\vert\tau)&=-\mathrm{i}y^{\frac{1}{2}}q^{\frac{1}{8}}\stackrel[n=1]{\infty}{\prod}(1-q^{n})(1-yq^{n})(1-y^{-1}q^{n-1}), \nonumber \\
\vartheta_{2}(z\vert\tau)&=y^{\frac{1}{2}}q^{\frac{1}{8}}\stackrel[n=1]{\infty}{\prod}(1-q^{n})(1+yq^{n})(1+y^{-1}q^{n-1}), \nonumber \\
\vartheta_{3}(z\vert\tau)&=\stackrel[n=1]{\infty}{\prod}(1-q^{n})(1+yq^{n-1/2})(1+y^{-1}q^{n-1/2}), \nonumber \\
\vartheta_{4}(z\vert\tau)&=\stackrel[n=1]{\infty}{\prod}(1-q^{n})(1-yq^{n-1/2})(1-y^{-1}q^{n-1/2}).
\end{align}

Jacobi's theta functions at $z=0$ are termed standard Jacobi's theta functions
\begin{equation}
\vartheta_{\nu}(\tau)\equiv\vartheta_{\nu}(0\vert\tau)
\end{equation}
with $\nu=1,2,3,4$. Hence we have
\begin{align}
\vartheta_{2}(\tau)&=\underset{n\in\mathbb{Z}}{\sum}q^{(n+1/2)^{2}/2}, \nonumber \\
\vartheta_{3}(\tau)&=\underset{n\in\mathbb{Z}}{\sum}q^{n^{2}/2}, \nonumber \\
\vartheta_{4}(\tau)&=\underset{n\in\mathbb{Z}}{\sum}(-1)^{n}q^{n^{2}/2},
\end{align}
or, equivalently,
\begin{align}
\vartheta_{2}(\tau)&=2q^{1/8}\stackrel[n=1]{\infty}{\prod}(1-q^{n})(1+q^{n})^{2}, \nonumber \\
\vartheta_{3}(\tau)&=\stackrel[n=1]{\infty}{\prod}(1-q^{n})(1+q^{n-1/2})^{2}, \nonumber \\
\vartheta_{4}(\tau)&=\stackrel[n=1]{\infty}{\prod}(1-q^{n})(1-q^{n-1/2})^{2}.
\end{align}
Note that $\vartheta_{1}(\tau)=0$, as can easily be seen from its infinite product form.

The (generalized) Weierstrass functions are defined in terms of Jacobi's theta functions:
\begin{equation}
\mathscr{P}_{\nu}(z_{i}-z_{j}\vert\tau)=\frac{\vartheta_{\nu}(z_{i}-z_{j}\vert\tau)\partial_{z}\vartheta_{1}(z\vert\tau)\vert_{z=0}}{\vartheta_{\nu}(\tau)\vartheta_{1}(z_{i}-z_{j}\vert\tau)}
\label{eq:Weierstrass}
\end{equation}
with $\nu=2,3,4$. When $z_{i}-z_{j}\rightarrow0$, these functions behave as $1/(z_{i}-z_{j})$, i.e., the two-point correlator $\langle\chi(z_{i})\chi(z_{j})\rangle$ on the plane is indeed recovered at short distance.

\subsection{Periodic properties}

Using Eqs.~\eqref{Jacobi_theta}, one can readily show that under the translations $z\rightarrow z+1$ and $z\rightarrow z+\tau$, Jacobi's theta functions transform as
\begin{align}
\label{translation_begin}
\vartheta_{1}(z+1\vert\tau)&=-\vartheta_{1}(z\vert\tau), \nonumber \\
\vartheta_{2}(z+1\vert\tau)&=-\vartheta_{2}(z\vert\tau), \nonumber \\
\vartheta_{3}(z+1\vert\tau)&=\vartheta_{3}(z\vert\tau), \nonumber \\
\vartheta_{4}(z+1\vert\tau)&=\vartheta_{4}(z\vert\tau),
\end{align}
and
\begin{align}
\vartheta_{1}(z+\tau\vert\tau)&=-y^{-1}q^{-1/2}\vartheta_{1}(z\vert\tau), \nonumber \\
\vartheta_{2}(z+\tau\vert\tau)&=y^{-1}q^{-1/2}\vartheta_{2}(z\vert\tau), \nonumber \\
\vartheta_{3}(z+\tau\vert\tau)&=y^{-1}q^{-1/2}\vartheta_{3}(z\vert\tau), \nonumber \\
\vartheta_{4}(z+\tau\vert\tau)&=-y^{-1}q^{-1/2}\vartheta_{4}(z\vert\tau),
\label{translation_end}
\end{align}
respectively.

Based on the translational properties of Jacobi's theta functions [Eqs.~(\ref{translation_begin}) and~(\ref{translation_end})], one can easily see that the (generalized) Weierstrass functions satisfy the periodicity conditions,
\begin{align}
\mathscr{P}_{2}(z+1\vert\tau)&=\mathscr{P}_{2}(z\vert\tau), \nonumber \\
\mathscr{P}_{3}(z+1\vert\tau)&=-\mathscr{P}_{3}(z\vert\tau), \nonumber \\
\mathscr{P}_{4}(z+1\vert\tau)&=-\mathscr{P}_{4}(z\vert\tau),
\label{eqs:Weierstrass_periodicity1}
\end{align}
and
\begin{align}
\mathscr{P}_{2}(z+\tau\vert\tau)&=-\mathscr{P}_{2}(z\vert\tau), \nonumber \\
\mathscr{P}_{3}(z+\tau\vert\tau)&=-\mathscr{P}_{3}(z\vert\tau), \nonumber \\
\mathscr{P}_{4}(z+\tau\vert\tau)&=\mathscr{P}_{4}(z\vert\tau),
\label{eqs:Weierstrass_periodicity2}
\end{align}
as the two-point correlators $\langle\chi(z_{i})\chi(z_{j})\rangle_{\nu}$ ($\nu=2,3,4$) on the torus should obey.

\subsection{Modular properties}

From the definition of Jacobi's theta functions~\eqref{Jacobi_theta}, one can readily verify that under the modular T transformation ($\tau\rightarrow\tau+1$) they behave as
\begin{align}
\vartheta_{1}(z\vert\tau+1)&=\mathrm{e}^{\frac{\mathrm{i}\pi}{4}}\vartheta_{1}(z\vert\tau), \nonumber \\
\vartheta_{2}(z\vert\tau+1)&=\mathrm{e}^{\frac{\mathrm{i}\pi}{4}}\vartheta_{2}(z\vert\tau), \nonumber \\
\vartheta_{3}(z\vert\tau+1)&=\vartheta_{4}(z\vert\tau), \nonumber \\
\vartheta_{4}(z\vert\tau+1)&=\vartheta_{3}(z\vert\tau).
\label{eqs:theta_T}
\end{align}
Using Eqs.~\eqref{eqs:theta_T} and the definition of $\mathscr{P}_{\nu}(z_{i}-z_{j}\vert\tau)$ in Eq.~(\ref{eq:Weierstrass}) (with $\tau=\mathrm{i}$ relevant for our setup),
it is easy to derive that
\begin{align}
\mathscr{P}_{2}(z_{i}-z_{j}\vert\mathrm{i}+1)&=\mathscr{P}_{2}(z_{i}-z_{j}\vert\mathrm{i}), \nonumber \\
\mathscr{P}_{3}(z_{i}-z_{j}\vert\mathrm{i}+1)&=\mathscr{P}_{4}(z_{i}-z_{j}\vert\mathrm{i}), \nonumber \\
\mathscr{P}_{4}(z_{i}-z_{j}\vert\mathrm{i}+1)&=\mathscr{P}_{3}(z_{i}-z_{j}\vert\mathrm{i}).
\label{eqs:Weierstrass_T}
\end{align}

The derivation of transforming properties of Jacobi's theta functions under the modular S transformation ($\tau\rightarrow-1/\tau, z\rightarrow z/\tau$) is more involved. To this end, we invoke the Poisson resummation formula~\cite{francesco1997,blumenhagen2009}
\begin{equation}
\underset{n\in\mathbb{Z}}{\sum}\exp\left(-\pi an^{2}+bn\right)=\frac{1}{\sqrt{a}}\underset{k\in\mathbb{Z}}{\sum}\exp\left[-\frac{\pi}{a}\left(k+\frac{b}{2\pi\mathrm{i}}\right)^{2}\right].
\end{equation}
With the definition of $\vartheta_{1}(z\vert\tau)$ and $a=\mathrm{i}/\tau$
and $b=-\mathrm{i}\pi(1-2z/\tau-1/\tau)$ in the above formula, we obtain
\begin{equation}
\vartheta_{1}(\tfrac{z}{\tau}\vert\tfrac{-1}{\tau})=-\mathrm{i}\mathrm{e}^{\mathrm{i}\pi\frac{z^{2}}{\tau}}(-\mathrm{i}\tau)^{\frac{1}{2}}\vartheta_{1}(z\vert\tau).
\label{eq:theta_S_1}
\end{equation}
Similarly, choosing $a=\mathrm{i}/\tau$ and $b=\mathrm{i}\pi(2z/\tau-1/\tau)$,
we obtain
\begin{equation}
\vartheta_{2}(\tfrac{z}{\tau}\vert\tfrac{-1}{\tau})=\mathrm{e}^{\mathrm{i}\pi\frac{z^{2}}{\tau}}(-\mathrm{i}\tau)^{\frac{1}{2}}\vartheta_{4}(z\vert\tau);
\label{eq:theta_S_2}
\end{equation}
choosing $a=\mathrm{i}/\tau$ and $b=2\mathrm{i}\pi z/\tau$, we obtain
\begin{equation}
\vartheta_{3}(\tfrac{z}{\tau}\vert\tfrac{-1}{\tau})=\mathrm{e}^{\mathrm{i}\pi\frac{z^{2}}{\tau}}(-\mathrm{i}\tau)^{\frac{1}{2}}\vartheta_{3}(z\vert\tau);
\label{eq:theta_S_3}
\end{equation}
and choosing $a=\mathrm{i}/\tau$ and $b=\mathrm{i}\pi(2z/\tau-1)$, we obtain
\begin{equation}
\vartheta_{4}(\tfrac{z}{\tau}\vert\tfrac{-1}{\tau})=\mathrm{e}^{\mathrm{i}\pi\frac{z^{2}}{\tau}}(-\mathrm{i}\tau)^{\frac{1}{2}}\vartheta_{2}(z\vert\tau).
\label{eq:theta_S_4}
\end{equation}
Using Eqs.~(\ref{eq:theta_S_1})--(\ref{eq:theta_S_4}) and the definition
of $\mathscr{P}_{\nu}(z_{i}-z_{j}\vert\tau)$ in Eq.~(\ref{eq:Weierstrass}) (with $\tau=-1/\tau=\mathrm{i}$ relevant for our setup), we obtain the change of Weierstrass functions under the modular S transformation,
\begin{align}
\mathscr{P}_{2}(-\mathrm{i}(z_{i}-z_{j})\vert\mathrm{i})&=\mathrm{i}\mathscr{P}_{4}(z_{i}-z_{j}\vert\mathrm{i}), \nonumber \\
\mathscr{P}_{3}(-\mathrm{i}(z_{i}-z_{j})\vert\mathrm{i})&=\mathrm{i}\mathscr{P}_{3}(z_{i}-z_{j}\vert\mathrm{i}), \nonumber \\
\mathscr{P}_{4}(-\mathrm{i}(z_{i}-z_{j})\vert\mathrm{i})&=\mathrm{i}\mathscr{P}_{2}(z_{i}-z_{j}\vert\mathrm{i}).
\label{eqs:Weierstrass_S}
\end{align}

We also note in passing that Dedekind's eta function~(\ref{eq:dedekind_eta}) transforms under the modular T and S transformations as
\begin{equation}
\eta(\tau+1)=\mathrm{e}^{\frac{\mathrm{i}\pi}{12}}\eta(\tau)
\label{eq:eta_T}
\end{equation}
and
\begin{equation}
\eta(-1/\tau)=(-\mathrm{i}\tau)^{\frac{1}{2}}\eta(\tau),
\label{eq:eta_S}
\end{equation}
respectively.

\section{``Jastrow times Pfaffian" form in Cartan basis}

\label{sec:cartan}

In this Appendix, we define the Cartan basis and show that the wave functions in Eq.~(\ref{eq:RVB}) take a ``Jastrow times Pfaffian" form in this basis.

The Cartan basis is defined by $\vert m\rangle=d_{m}^{\dagger}\vert0\rangle$ with $m=1,0,-1$. (Again, we have temporarily suppressed the site index.) Here $\vert m=0 \rangle$ should not be confused with the parton vacuum $\vert0\rangle$. We shall also frequently use $\vert \pm \rangle$ as the shorthand notation for $\vert m=\pm 1 \rangle$. The new parton operators $d_m^{\dag}$ are unitarily related to $c_a^{\dag}$ [see Eq.~(\ref{eq:localstate})] via
\begin{eqnarray}
d_{\pm}^{\dagger}&=&-\frac{1}{\sqrt{2}}(\pm c_{1}^{\dagger}+\mathrm{i}c_{2}^{\dagger}), \nonumber \\
d_{0}^{\dagger}&=&c_{3}^{\dagger}.
\end{eqnarray}
Thus, the Cartan basis is related to the ``color'' basis as
$\vert \pm \rangle=-(\pm\vert 1\rangle + \mathrm{i} \vert 2 \rangle)/\sqrt{2}$ and
$\vert m=0 \rangle=\vert 3\rangle$. In the Cartan basis, the $\mathfrak{so}(3)$ generator $J_3$ [see Eq.~(\ref{eq:generator})] is diagonalized, $J_3\vert m \rangle = m\vert m \rangle$, i.e., $m$ is the magnetic quantum number for the spin-1 system.

In the Cartan basis, the $\mathrm{SO}(3)$ valence bond
operator is rewritten as
\begin{equation}
\stackrel[a=1]{3}{\sum}c_{i,a}^{\dagger}c_{j,a}^{\dagger}=-(d_{i,+}^{\dagger}d_{j,-}^{\dagger}-d_{i,0}^{\dagger}d_{j,0}^{\dagger}+d_{i,-}^{\dagger}d_{j,+}^{\dagger}).
\end{equation}
Hence the RVB states in Eq.~(\ref{eq:RVB}) are rewritten as
\begin{eqnarray}
\vert\Psi_{\nu}\rangle&=&P_{\textrm{G}}\exp\left(\underset{i<i'}{\sum}\mathscr{P}_{\nu}(z_i-z_{i'}\vert\tau)d_{i,0}^{\dagger}d_{i',0}^{\dagger}\right) \nonumber \\
&\phantom{=}&\times\exp\left(-\underset{j\neq l}{\sum}\mathscr{P}_{\nu}(z_j-z_l\vert\tau)d_{j,-}^{\dagger}d_{l,+}^{\dagger}\right)\vert0\rangle.
\end{eqnarray}
Expanding two exponentials and performing the Gutzwiller projection, we obtain
\begin{eqnarray}
\vert\Psi_{\nu}\rangle&=& \stackrel[N_1=0]{N/2}{\sum} \stackrel[N_{0}=0]{N}{\sum} \delta_{2N_1+N_0,N} \underset{i_{1}<\cdots<i_{N_{0}}}{\sum} \underset{j_1<\cdots<j_{N_{1}}}{\sum}\underset{l_1<\cdots<l_{N_{1}}}{\sum} \nonumber \\
&\phantom{=}& \times \mathrm{Pf}_0\left[\mathscr{P}_{\nu}(z_i-z_{i'}\vert\tau)\right] \mathrm{det}_{\mp}\left[-\mathscr{P}_{\nu}(z_{j}-z_{l}\vert\tau)\right] \nonumber \\
&\phantom{=}&\times d_{i_{1},0}^{\dagger}\cdots d_{i_{N_{0}},0}^{\dagger} d_{j_{1},-}^{\dagger}d_{l_{1},+}^{\dagger}\cdots d_{j_{N_{1}},-}^{\dagger}d_{l_{N_{1}},+}^{\dagger}\vert0\rangle
\label{eq:RVB_Cartan}
\end{eqnarray}
with site indices $i_1<\cdots<i_{N_0}$ ($j_1<\cdots<j_{N_1}$ and $l_1<\cdots<l_{N_1}$) denoting the positions of $\vert m=0 \rangle$ ($\vert m=-1 \rangle$ and $\vert m=+1 \rangle$). Note that these site indices must all be distinct due to the Gutzwiller projection. Furthermore, the Pfaffian (determinant) factor is associated with the $\vert m=0 \rangle$ ($\vert m = \mp 1 \rangle$) states.

The determinant in Eq.~(\ref{eq:RVB_Cartan}) can be simplified by using the Cauchy determinant formula on the torus~\cite{francesco1997},
\begin{eqnarray}
&\phantom{=}& \mathrm{det}_{\mp}\left[\mathscr{P}_{\nu}(z_{j_p}-z_{l_q}\vert\tau)\right]_{1\leq p,q\leq N_1} \nonumber \\
&=&(-1)^{\frac{N_{1}(N_{1}-1)}{2}} \left(\partial_{z}\vartheta_{1}(z\vert\tau)\vert_{z=0}\right)^{N_{1}} \frac{\vartheta_{\nu}(\sum_p (z_{j_p}-z_{l_p})\vert\tau)}{\vartheta_{\nu}(0\vert\tau)} \nonumber \\
&\phantom{=}& \times\frac{\prod_{p<q}\vartheta_{1}(z_{j_p}-z_{j_q}\vert\tau)\vartheta_{1}(z_{l_p}-z_{l_q}\vert\tau)}{\prod_{p,q}\vartheta_{1}(z_{j_p}-z_{l_q}\vert\tau)}.
\label{eq:Cauchy_determinant_formula}
\end{eqnarray}
Substituting Eq.~\eqref{eq:Cauchy_determinant_formula} into Eq.~(\ref{eq:RVB_Cartan}) and noticing that the sign factor $(-1)^{\frac{N_{1}(N_{1}-1)}{2}}$ is cancelled by bringing the partons from
the ordering $d_{j_1,-}^{\dag}d_{l_1,+}^{\dag}\cdots d_{j_{N_{1}},-}^{\dag}d_{l_{N_{1}},+}^{\dag}$
to $d_{j_1,-}^{\dag}\cdots d_{j_{N_{1}},-}^{\dag} d_{l_1,+}^{\dag}\cdots d_{l_{N_{1}},+}^{\dag}$, we obtain
\begin{eqnarray}
\vert\Psi_{\nu}\rangle&=& \stackrel[N_1=0]{N/2}{\sum} \stackrel[N_{0}=0]{N}{\sum} \delta_{2N_1+N_0,N} \underset{i_{1}<\cdots<i_{N_{0}}}{\sum} \underset{j_1<\cdots<j_{N_{1}}}{\sum}\underset{l_1<\cdots<l_{N_{1}}}{\sum} \nonumber \\
&\phantom{=}& \times \mathrm{Pf}_0\left[\mathscr{P}_{\nu}(z_i-z_{i'}\vert\tau)\right] \frac{\vartheta_{\nu}(\sum_p (z_{j_p}-z_{l_p})\vert\tau)}{\vartheta_{\nu}(0\vert\tau)} \nonumber \\
&\phantom{=}& \times (-1)^{N_{1}}\frac{\prod_{p<q}E(z_{j_p}-z_{j_q}\vert\tau)E(z_{l_p}-z_{l_q}\vert\tau)}{\prod_{p,q}E(z_{j_p}-z_{l_q}\vert\tau)} \nonumber \\
&\phantom{=}& \times
d_{i_{1},0}^{\dag}\cdots d_{i_{N_{0}},0}^{\dag}
d_{j_1,-}^{\dag}\cdots d_{j_{N_{1}},-}^{\dag} d_{l_1,+}^{\dag}\cdots d_{l_{N_{1}},+}^{\dag}\vert0\rangle \nonumber
\end{eqnarray}
with the definition
\begin{equation}
E(z\vert\tau)=\frac{\vartheta_{1}(z\vert\tau)}{\partial_{z}\vartheta_{1}(z\vert\tau)\vert_{z=0}}.
\end{equation}

Finally, by rearranging the order of the partons, the wave function is written in a ``Jastrow times Pfaffian" form in the Cartan basis,
\begin{equation}
\vert\Psi_{\nu}\rangle=\sum_{m_{1},\ldots,m_{N}} \Psi_{\nu}(m_{1},\ldots,m_{N}) \vert m_{1},\ldots,m_{N}\rangle,
\end{equation}
where the basis is given by
\begin{equation}
\vert m_{1},\ldots,m_{N}\rangle=d_{1,m_{1}}^{\dagger}\cdots d_{N,m_{N}}^{\dagger}\vert0\rangle,
\end{equation} and the superposition coefficient is, after dropping an unimportant overall factor, written as
\begin{eqnarray}
\Psi_{\nu}(\{m\})&=& \rho_{m}\mathrm{Pf}_{0}\left[\mathscr{P}_{\nu}(z_{i}-z_{j}\vert\tau)\right]
\frac{\vartheta_{\nu}(\sum_{j=1}^{N} m_{j}z_{j}\vert\tau)}{\vartheta_{\nu}(0\vert\tau)}\nonumber \\
&\phantom{=}& \times \prod_{1\leq i<j\leq N}\left[E(z_{j}-z_{i}\vert\tau)\right]^{m_{i}m_{j}}
\label{eq:jastrow_pfaffian}
\end{eqnarray}
with $\rho_{m}=\prod_{j:\textrm{even}}(-1)^{m_j}$ if $\sum_{j=1}^{N}m_j = 0$ and $\rho_{m}=0$ otherwise.

\section{Proof of translational invariance}

\label{sec:translational-invariance}

In this Appendix, we prove that the RVB states~\eqref{eq:RVB} defined on the square lattice (see Sec.~\ref{sec:modular}) are invariant under translations along the two global loops of the torus. The coordinates of the lattice sites follow the notation introduced in Sec.~\ref{sec:modular}.

The translation by one lattice spacing along the $x$ direction acts on the fermionic operators $c_{j,a}^{\dagger}\equiv c_{(x_{j},y_{j}),a}^{\dagger}$ as
\begin{equation}
T_x c_{j,a}^{\dagger}T_x^{-1}=\begin{cases}
\begin{array}{c}
\mathrm{e}^{\mathrm{i}\theta_{x\nu}}c_{(0,y_{j}),a}^{\dagger} \\
c_{(x_{j}+1,y_{j}),a}^{\dagger}
\end{array} & \begin{array}{c}
x_{j}=L-1 \\
\textrm{otherwise}
\end{array} \end{cases},
\label{eq:Tx_operator}
\end{equation}
where the phase angle $\theta_{x\nu}$ is introduced to represent the (periodic or antiperiodic) boundary condition along the $x$ direction, i.e., $\theta_{x\nu}=0$ for $\nu=2$ and $\pi$ for $\nu=3,4$. Similarly, the translation by one lattice spacing along the $y$ direction acts on the fermionic operators as
\begin{equation}
T_y c_{j,a}^{\dagger}T_y^{-1}=\begin{cases}
\begin{array}{c}
\mathrm{e}^{\mathrm{i}\theta_{y\nu}}c_{(x_{j},0),a}^{\dagger} \\
c_{(x_{j},y_{j}+1),a}^{\dagger}
\end{array} & \begin{array}{c}
y_{j}=L-1 \\
\textrm{otherwise}
\end{array}\end{cases},
\label{eq:Ty_operator}
\end{equation}
where $\theta_{y\nu}=0$ for $\nu=4$ and $\pi$ for $\nu=2,3$.

By formally setting $\mathscr{P}_{\nu}(0\vert\tau)=0$, the RVB states in Eq.~(\ref{eq:RVB}) are written as
\begin{equation}
\vert\Psi_{\nu}\rangle=P_{\textrm{G}}\exp\left(\frac{1}{2}\sum_{i,j=1}^{N}\mathscr{P}_{\nu}(z_{i}-z_{j}\vert\tau)\sum_{a=1}^{3}c_{i,a}^{\dagger}c_{j,a}^{\dagger}\right)\vert0\rangle.
\end{equation}
By using $T_{x}P_{\textrm{G}}T_{x}^{-1}=P_{\textrm{G}}$ and $T_{x}\vert0\rangle=\vert0\rangle$, the action of $T_x$ upon $\vert\Psi_{\nu}\rangle$ is expressed as
\begin{eqnarray}
&\phantom{=}& T_x \vert\Psi_{\nu}\rangle \nonumber \\
&=& P_{\textrm{G}}T_x\exp\left(\frac{1}{2}\sum_{i,j=1}^{N}\mathscr{P}_{\nu}(z_{i}-z_{j}\vert\tau)\sum_{a=1}^{3}c_{i,a}^{\dagger}c_{j,a}^{\dagger}\right)T_x^{-1} \vert0\rangle   \nonumber \\
&=& P_{\textrm{G}}\exp\left(\sum_{a=1}^{3}\frac{1}{2}\sum_{i,j=1}^{N}\mathscr{P}_{\nu}(z_{i}-z_{j}\vert\tau) T_x c_{i,a}^{\dagger}c_{j,a}^{\dagger}T_x^{-1}\right) \vert0\rangle, \nonumber
\label{eq:Tx_action}
\end{eqnarray}
where the action of $T_x$ on the VB operators can be calculated separately in four cases: case 1, $x_i \neq L-1$ and $x_j \neq L-1$; case 2, $x_i \neq L-1$ and $x_j = L-1$; case 3, $x_i = L-1$ and $x_j \neq L-1$; and case 4, $x_i = x_j = L-1$. By using Eqs.~(\ref{eq:Tx_operator}) and (\ref{eq:Ty_operator}), we have
\begin{eqnarray}
T_x c_{i,a}^{\dagger}c_{j,a}^{\dagger}T_x^{-1}=\begin{cases}
\begin{array}{c}
c_{(x_i+1,y_i),a}^{\dagger}c_{(x_j+1,y_j),a}^{\dagger} \\
\mathrm{e}^{\mathrm{i}\theta_{x\nu}}c_{(x_i+1,y_i),a}^{\dagger}c_{(0,y_j),a}^{\dagger} \\
\mathrm{e}^{\mathrm{i}\theta_{x\nu}} c_{(0,y_i),a}^{\dagger}c_{(x_j+1,y_j),a}^{\dagger} \\
\mathrm{e}^{2\mathrm{i}\theta_{x\nu}} c_{(0,y_i),a}^{\dagger}c_{(0,y_j),a}^{\dagger}
\end{array} & \begin{array}{c}
\textrm{case 1} \\
\textrm{case 2} \\
\textrm{case 3} \\
\textrm{case 4}
\end{array} \end{cases} \nonumber.
\end{eqnarray}
However, the phase factors arising above are cancelled. Let us take case 2 as an example,
\begin{eqnarray}
&\phantom{=}& \sum_{x_i=0}^{L-2}\sum_{y_i,y_j=0}^{L-1} \mathscr{P}_{\nu}(z_i-z_j\vert\tau) T_x c_{(x_i,y_i),a}^{\dagger}c_{(L-1,y_j),a}^{\dagger} T_x^{-1} \nonumber \\
&=& \sum_{x_i=0}^{L-2}\sum_{y_i,y_j=0}^{L-1} \mathscr{P}_{\nu}( \tfrac{1}{L}[x_i - (L-1) + \mathrm{i}(y_i-y_j)] \vert\tau)  \nonumber \\
&\phantom{=}& \times \mathrm{e}^{\mathrm{i}\theta_{x\nu}} c_{(x_i+1,y_i),a}^{\dagger}c_{(0,y_j),a}^{\dagger} \nonumber \\
&=& \sum_{x'_i=1}^{L-1}\sum_{y_i,y_j=0}^{L-1} \mathscr{P}_{\nu}( \tfrac{1}{L}[x'_i + \mathrm{i}(y_i-y_j)]-1 \vert\tau)   \nonumber \\
&\phantom{=}& \times \mathrm{e}^{\mathrm{i}\theta_{x\nu}} c_{(x'_i,y_i),a}^{\dagger}c_{(0,y_j),a}^{\dagger} \nonumber \\
&=& \sum_{x'_i=1}^{L-1}\sum_{y_i,y_j=0}^{L-1} \mathscr{P}_{\nu}( \tfrac{1}{L}[x'_i + \mathrm{i}(y_i-y_j)] \vert\tau) c_{(x'_i,y_i),a}^{\dagger}c_{(0,y_j),a}^{\dagger}, \nonumber
\end{eqnarray}
where we have used an identity [see Eq.~(\ref{eqs:Weierstrass_periodicity1})]
\begin{equation}
\mathscr{P}_{\nu}(z\pm 1 \vert\tau)=\mathrm{e}^{\mathrm{i}\theta_{x\nu}}\mathscr{P}_{\nu}(z\vert\tau).
\end{equation}
After collecting the terms from all four cases, it is easy to see that the sum of all VB operators is invariant under the action of $T_x$. Thus, we obtain $T_x\vert\Psi_{\nu}\rangle = \vert\Psi_{\nu}\rangle$, which proves the translational invariance of $\vert\Psi_{\nu}\rangle$ along the $x$ direction.

The translational invariance of $\vert\Psi_{\nu}\rangle$ along the $y$ direction can be similarly proved with the help of an identity  [see Eq.~(\ref{eqs:Weierstrass_periodicity2})]
\begin{equation}
\mathscr{P}_{\nu}(z\pm\tau\vert\tau)=\mathrm{e}^{\mathrm{i}\theta_{y\nu}}\mathscr{P}_{\nu}(z\vert\tau).
\end{equation}
This completes the proof that the RVB states on the square lattice are translationally invariant along the two global loops of the torus.

\bibliography{moore_read}

\end{document}